\documentclass[twocolumn,showpacs,preprintnumbers,amsmath,amssymb]{revtex4}
\usepackage{graphicx}% Include figure files
\usepackage{dcolumn}% Align table columns on decimal point
\usepackage{bm}% bold math
\usepackage[usenames]{color}

\begin{document}

\title{On the non-convergence of the Wang-Landau algorithms with multiple random walkers}% Force line breaks with \\

\author{R. E. Belardinelli$^{\dag,\S}$}
\author{V. D. Pereyra$^\S$}
\affiliation{%
$^{\dag}$Instituto de F\'{\i}sica Aplicada (INFAP) - CONICET,$^\S$ Departamento de F\'{\i}sica, Universidad Nacional de San Luis, CONICET, Chacabuco 917, 5700
San Luis, Argentina
%\textbackslash
}%

\begin{abstract}
This paper discusses some convergence properties in the entropic sampling Monte Carlo methods with multiple random walkers, particularly in the Wang-Landau (WL) and $1/t$ algorithms. The classical algorithms are modified by the use of $m$ independent random walkers in the energy landscape to calculate the density of states (DOS). The Ising model is used to show the convergence properties in the calculation of the DOS, as well as the critical temperature, while the calculation of the  number $\pi$ by multiple dimensional integration is  used in the continuum approximation. In each case, the error is obtained separately for each walker at a fixed time, $t$; then, the average over $m$ walkers is performed. It is observed that the error goes as $1/\sqrt{m}$. However, if the number of walkers increases above a certain critical value $m>m_x$,  the error reaches a constant value (i.e. it saturates). This occurs for both algorithms; however, it is shown  that for a given system, the $1/t$ algorithm is more efficient and accurate than the similar version of the WL algorithm. It follows that it makes no sense to increase the number of walkers above a critical value $m_x$, since it does not reduces the error in the calculation. Therefore, the number of walkers does not guarantee convergence.
\end{abstract}

% PACS, the Physics and Astronomy
                             % Classification Scheme.
\maketitle

\section{INTRODUCTION}

The Wang-Landau (WL) algorithm is currently one of the most widely used variations of the Monte Carlo simulation method introduced in the last years [\onlinecite{Wang01a,Wang01b,Wang02}]. It belongs to the broader class of flat-histogram Monte Carlo simulations, aimed at obtaining an estimate of the density of states (DOS) $g(E)$ of a system with high accuracy ($g(E)$ represents the number of possible states or configurations with energy $E$).

Recent studies have proposed improvements and sophisticated implementations of the WL algorithm [\onlinecite{Day04,Zho05,Oka06,Lee06,Zho08,Chi09,Fyt10,Mal10,Than10,Dic11,Bro11,depablo03,Earl05,Morozov07,li07,Wus12}].

The convergence properties of the original WL formulation has been an issue of controversy. In fact, several studies show that the saturation of the final error persists (i. e., the difference between the simulation estimates for $g(E)$ and the exact values) regardless of the simulation effort employed. This problem was first pointed out by Q. Yang and J. J. de Pablo in reference \cite{depablo03}. Several authors [\onlinecite{Zho05,Lee06,Zho08,Earl05,Morozov07}] have also analyzed the WL convergence. In particular, Zhou and Bhatt \cite{Zho05} presented an argument for its convergence.

It is well known that the exponential decrease of the modification factor  $F=\ln f$ (which is defined below) with the number of iterations, is the reason for the saturation of the error in the original WL algorithm, so that in the final sampling stages, the error to estimate $g(E)$ is essentially constant. To overcome this limitation, a new version of the WL algorithm has been introduced in references [\onlinecite{Bela07a,Bela07b,Bela08,Bela14}], in which the modification factor  is scaled down as $1/t$ instead of exponentially. 

The $1/t$ algorithm has been successfully applied to several statistical systems [\onlinecite{Fy08,Fy09,Oje09,Mal09,Oje10,Fy11,Swe11,Kom12,Ing13,Jank16}]. 

Very recently a new version of the WL algorithm, the Stochastic Approximation Monte Carlo [\onlinecite{Wer14,Wer15}], which uses the $1/t$ strategy has been successfully applied to semi-flexible polymers chains.

The convergence of the $1/t$ algorithm has been discussed in previous work[\onlinecite{Bela07b,Bela14,Zho08}]. In fact, it has been analytically demonstrated \cite{Bela07b} that the entropy $S(E,t)=\ln g(E,t)$ can be expressed as a series in which $F(t)$ is the kernel; in those algorithms where $F_k=F_{k-1}/l$ (with any value of $l>1$), the resulting series converges to a finite value, and then, the error reaches a constant value (saturates in time). On the contrary, in those algorithms where the modification factor depends on time as $F(t)=t^{-\gamma}$ with $\gamma\leq 1$ (the optimum choice is $\gamma=1$), the series is divergent and the calculated density of states approaches asymptotically to the exact values as $\approx t^{-\gamma/2}$.

Recently, the tomographic sampling method was modified by the use of the $1/t$ scheme. The tomographic algorithm was originally implemented using, in effect, a modification factor  $F =\ln f $ that does not change with time \cite{Di11}. It is shown that convergence is improved by using $F \sim 1/t$ in this method as well \cite{Bela14}. Besides that, the authors have demonstrated that there is convergence in the case that $0 < \gamma \le 1$, by using on analytical argument applied to the simple two-state model.

Nevertheless, numerical studies show that the error in the $1/t$ algorithm decays as $1/\sqrt{t}$; and to our knowledge, this has not been improved upon.

Therefore, whenever the modification factor $ F $ decreases exponentially with the number of iterations, the algorithm does not converge, regardless of any modification of the WL algorithm. That is, the error in calculating the DOS approaches a constant value (i.e., reaches saturation), as it has been analytically proved in references [\onlinecite{Bela07b,Bela14}]. 

A general comment on the use of WL algorithm: despite the problems of convergence, it is nowadays well known that the WL method works very well for getting a first approximation of the density of states and then use it as an ingredient of a controlled numerical scheme (any type of multi-histogram method). However, far from a criticism, any new contribution that helps to understand the behavior of the method and solve the problem of convergence should be considered.

In references [\onlinecite{Vog13,Vog14,Shi14,Li14}], D. P. Landau and co-workers introduced a massive parallel WL sampling based on the replica-exchange framework for Monte Carlo simulations. They introduced $m$ random walkers in a energy sub-window.
They emphasize that the estimated density of states converges to the true one with an increasing number of iterations, and the simulation is terminated when the modification factor reaches a minimal value $f_{final}$. They demonstrated the advantages and general applicability of the method for the simulation of complex systems. They also showed that this algorithm is extremely efficient and that its parallel implementation is straightforward. This practice reduces the error during the simulation with $ 1/\sqrt {m}$, where $m$ is the number of independent walkers in the energy sub-window.

A similar strategy to reduce the error with the number of walkers is implemented in references [\onlinecite{Ca12,Ca14}]. Parallel implementation of other multi-histogram methods are introduced in reference \cite{Jan13}.

Although the method has been implemented in a massive parallel sampling in systems with multiple windows, it is easy to apply to a system with a single window. 

Thus, even when the error saturates for a single random walk, the average of $m$ random walkers seems to converge to the exact value, i.e., the error seems to depend on $ 1/\sqrt {m}$. 

In this paper, the validity of this assumption is discussed. For that, a simple implementation of the algorithms is performed to calculate the DOS and other observables such as the critical temperature in the Ising model and the number $\pi$ by numerical integration. The Ising model and numerical integration is used as a laboratory test for different reasons, i.e.:  i) the DOS is known, at least, for small systems, ii) the observable can be obtained with high precision, iii) if it does not converge for  trivial systems like the  mentioned above (which presents a relatively well behaved energy landscape), it seems unlikely that the converges in more complex system.
The remainder of this paper is organized as follows: in Section 1, the algorithms and different quantities are introduced, as well as the definition of the errors for the different models. In Section 2, the algorithms and their implementations are discussed. The results and the conclusions are presented in Section 3 and 4, respectively.

\begin{figure}
\includegraphics[scale=0.75]{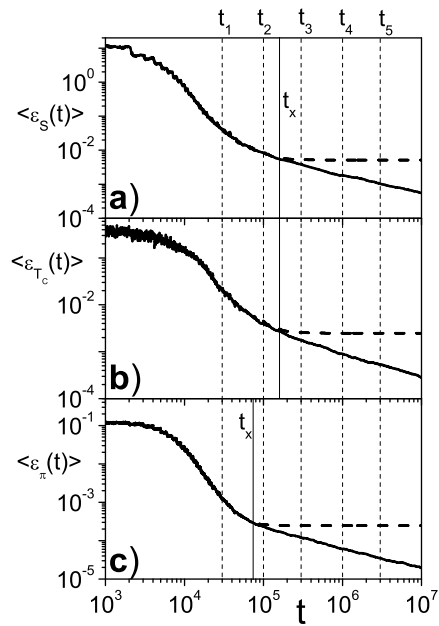}% Here is how to import EPS art
\caption{\label{fig:epsart} Behavior of the error as a function of time using the WL algorithm (long dashed line) and the $1/t$ algorithm (solid line) for a single walker. The data correspond to a) the DOS and b) the critical temperature, obtained from the peak location of the specific heat for a two-dimensional Ising Model; and c) the calculation of the number $\pi$ using multidimensional integrations. The critical time $t_x$, corresponding to the saturation of the error using the WL algorithm, is shown in figures (vertical solid line). The times $t_1, t_2, t_3, t_4$ and $t_5$ correspond to the times at which the algorithm is stopped to start the $m$ walkers, these are indicated with vertical dashed lines. The slope of the curves corresponding to the $1/t$ algorithm goes as $1/\sqrt{t}$. The data represent the average of 200 independent realizations ($p$=200).}
\end{figure}

\begin{figure}
\includegraphics[scale=0.75]{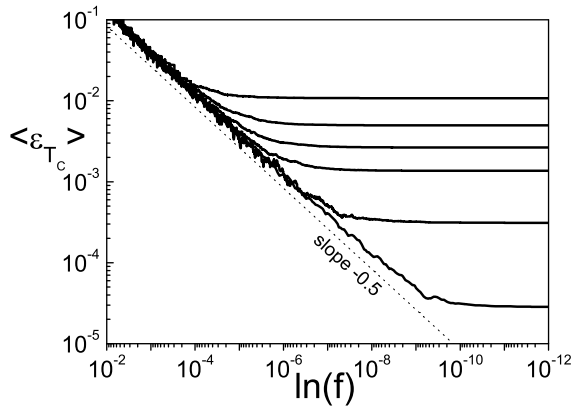}% Here is how to import EPS art
\caption{\label{fig:epsart} The error as a function of the flatness criteria calculated for the WL algorithm. The curves are shown in decreasing order with $50 \%$, $80 \%$, $90 \%$, $95 \%$, $99 \%$ and $99.9 \%$, respectively. The data correspond to the critical temperature, obtained from the peak location of the specific heat for a two-dimensional Ising Model with $L=8$ and it is the average of 200 independent realizations ($p$=200).}
\end{figure}

\section{ALGORITHMS AND THEIR IMPLEMENTATIONS}

The density of states in energy, $g(E)$, measures the energy degeneracy of the admissible states of a system, from which the partition function $Z$ can be calculated:
\begin{eqnarray}
Z(T)=\sum_{\rho} e^{-E[\rho]/k_BT}=\sum_{E}g(E) e^{-E/k_BT}
\end{eqnarray}
where $\rho$ stands for a state or configuration that the system can reside in; $k_B$ is the Boltzmann constant and $T$ is the temperature. The first sum runs over all possible states of the system, whereas the second sum runs over all possible total energies and it can only be calculated once $g(E)$ is known. While $g(E)$ is temperature independent and only depends on the definition of the Hamiltonian, Eq.(1) allows for the calculation of the temperature dependent $Z$ via the corresponding Boltzmann factors.
One also defines a dimensionless entropy $S(E) \equiv k_B\ln g(E)$.
An important consequence is the possibility of calculating the thermodynamic quantities at any temperature with the sole knowledge of $g(E)$. For example, the average energy $\langle E \rangle$:
\begin{eqnarray}
\langle E(T)\rangle = \frac{1}{Z}\sum_{E}g(E) E e^{-E/k_BT}
\end{eqnarray}
and the heat capacity $C_V$ can be calculated as:
\begin{eqnarray}
C_V(T)= \frac{\langle E^2 \rangle-(\langle E \rangle)^2}{k_BT^2}
\end{eqnarray}
These thermodynamic observables provide a measure to identify and locate phase transitions, and hence understand critical phenomena. 

The standard WL algorithm [\onlinecite{Wang01a,Wang01b,Wang02}] estimates the DOS using a
single random walker in an energy range $[E_{min}, E_{max}]$. During the simulation, trial moves are
accepted with a probability $P = min [1, g(E_{old})/g(E_{new})]$, where $E_{old} (E_{new})$ is the energy of the
original (proposed) configuration. The estimation of $g(E)$ is continuously adjusted and improved
using a modification factor $f$ (i.e. $g(E) \rightarrow  f \times g(E)$), which starts with $f_0 > 1$ and progressively
approaches unity as the simulation proceeds. A histogram, $H(E)$, keeps track of the number
of visits to each energy $E$ during a given iteration. When $H(E)$ is sufficiently $flat$ \cite{Flatt00}, the next
iteration begins with $H(E)$ reset to zero but keeping the estimate of $g(E)$ from the previous
iteration, and $f$ reduced by some predefined rule (e.g. $f \rightarrow \sqrt{f}$). The simulation ends when $f$ reaches a sufficiently small value $f_{stop}$, at which point the accuracy of $g(E)$ is proportional to $\sqrt{f_{stop}}$ for sufficiently flat $H(E)$.
The $1/t$ algorithm works as the original WL algorithm, but as soon as $F=  \ln (f) \leq 1/t$, $F \rightarrow 1/t$; thereafter, $F(t) = 1/t$ is updated at each event (here, $t$ is the Monte Carlo time defined as $t = n/N$, where $n$ is the number of
attempted changes of state, or steps, and $N$ is the energy range). In other words, for a characteristic time, the modification factor  $F$ goes from exponential to power decay. For more details see reference \cite{Bela14}.

To assess its applicability, feasibility, and performance,  the  $1/t$ and the WL $m$-random walkers are applied to the two-dimensional Ising model on square lattices, as well as the calculation of the number $\pi$ by numerical integration.

The two-dimensional Ising model on a square lattice with linear size $L=8$ and periodic boundary conditions, is used for the calculation of the DOS and the critical temperature. The size of the system is similar to the size of a single window in the replica exchange Wang Landau sampling [\onlinecite{Vog13,Vog14,Li14}]. Despite of the size, this is sufficient for the purposes of this study. However, in order to show the effect of the size on the behavior of the error, the study is also applied to a window of $N=300$ which belongs to larger system size $L=64$.

Monte Carlo multidimensional integration using WL and $1/t$ algorithms, are also implemented to calculate the number $\pi$ [\onlinecite{li07,Bela08,Ati15}].

Before discussing the results it is necessary to explain how proceed to calculate the errors of the quantities to be measured. In this paper, the error in the calculation of the DOS, as the number of walkers $ m $, and $ t $ time is defined as:

\begin{eqnarray}
\varepsilon_{S}(t,m)=\frac{1}{N}\sum_{E}\bigg |\frac{\overline{S_{E}(t,m)}-S_{E}^{exc}}{S_{E}^{exc}}\bigg |,
\end{eqnarray}
where $S_{E}^{exc}$ is the exact value of the DOS for the energy $E$. The average over the number of walkers $m$ is indicated by the top-line, which is given by:

\begin{eqnarray}
\overline{S_{E}(t,m)} =\frac{1}{m}\sum_{i}^{m}S_{E,i}(t).
\end{eqnarray}
Similarly, one proceeds with the error in the calculation of the observable $X(t,m)$:  
\begin{eqnarray}
\varepsilon_{X}(t,m)=\bigg |\frac{\overline{X(t,m)} - X^{exc}}{X^{exc}}\bigg |,
\end{eqnarray}
where
\begin{eqnarray}
\overline{X(t,m)}=\frac{1}{m}\sum_{i}^{m}X_i(t),
\end{eqnarray}
the corresponding standard deviation is:
\begin{eqnarray}
\sigma_{m}=\sqrt{\overline{X(t,m)^2}-(\overline{X(t,m))}^2}
\end{eqnarray}

From the above equations, the errors as a function of time for $ m $ walkers, can obtained. To smooth the curves, which usually present some noise, average over $p$ independent realizations is performed. The average is indicated by brackets and is defined as:
\begin{eqnarray}
\langle \varepsilon(t,m) \rangle=\frac{1}{p}\sum_{i}^{p}\varepsilon_i(t,m).
\end{eqnarray}
Similarly, for the mean value
\begin{eqnarray}
\langle \overline {X(t,m)} \rangle=\frac{1}{p}\sum_{i}^{p}\overline {X_i(t,m)},
\end{eqnarray}
and the standard deviation  
\begin{eqnarray}
\sigma_{p}=\sqrt{\langle \overline {X(t,m)}^2\rangle-\langle \overline{X(t,m)}\rangle ^2}.
\end{eqnarray}

\begin{figure}
\includegraphics[scale=0.75]{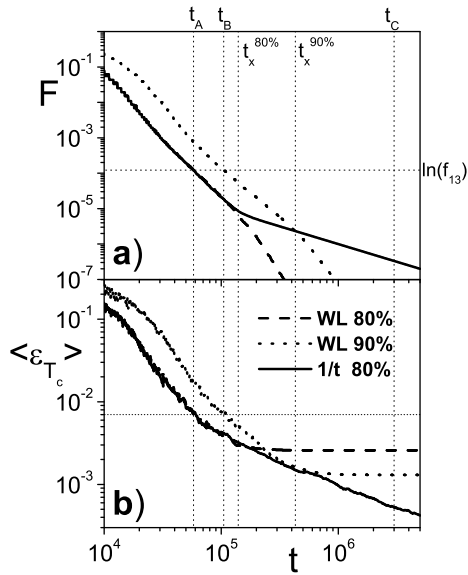}% Here is how to import EPS art
\caption{\label{fig:epsart} Behavior of (a) $F(t)=\ln f$; and (b) the error in the calculation of the critical temperature for a two-dimensional Ising Model  $\epsilon_{T_c}(t)$, calculated by using the $WL$ for $80 \%$ and $90 \%$-flatness criteria, and $1/t$ algorithms. Vertical solid lines represent the saturation times for the WL algorithm ($t_x^{80}\approx 140000$ MCS and $t_x^{90}\approx 430000$ MCS). The times $t_A \approx 58000$ MCS, $t_B \approx 105000$ MCS and $t_C\approx 3\times 10^6$ MCS as $F(t)=\ln f_{13}=1.2208\times 10^{-4}$ are described in the text. The data represent the average of 200 independent realizations ($p$=200).}
\end{figure}

\section{DISCUSSION}
Firstly, the convergence properties of the WL and the $1/t$ algorithms as a function of time $t$, for a single walker, are discussed. The error in the calculation of the DOS as a function of time, $t$, for a two-dimensional Ising model, with linear size $L$ is obtained from the above equations. Note that, for m = 1, the error in the DOS (Eq. (4)), and in the observable (Eq (6)), is in agreement with the definition given in references [\onlinecite{Bela07a,Bela07b,Bela08,Bela14}], and $\sigma_{m=1}=0$. In this case, the range of the energy is $N=L^2-1$. The exact density of states, as well as the exact critical temperature for the Ising model, with a given system size, are obtained by using the methodology developed by Beale in reference \cite{Beale96}. Similarly, one can obtain the error in the calculation of the number $\pi$, as discussed in reference [\onlinecite{li07,Bela08,Ati15}].

In Figure 1, different errors are shown as a function of time, using the WL and $1/t$ algorithms. As described in reference \cite{Bela14}, the $1/t$ algorithm presents two temporal regimes: i) the first stage of the simulation, where the  $ 1 / t $ algorithm coincides with the WL, and the error decreases sharply; and ii) the second stage, when $ F_i < 1/t $. Time $t=t_x$ separates the two regimes and coincides with the saturation time of the WL algorithm (where the error becomes constant). This time is shown in Figure 1 with a vertical solid line. The saturation time for a two-dimensional Ising Model with $L=8$, is $t_x \approx 140000$ MCS, using $80 \%$-flatness criterion. As expected, for Figures 1a and 1b, $t_x$ coincides; while the corresponding saturation time for the calculation of $\pi$ with WL algorithm is $t_x=74000$ MCS. On the other hand, the slope of the curves corresponding to the $1/t$ algorithm goes as $1/\sqrt{t}$. The times $t_1, t_2, t_3, t_4$ and $t_5$  are characteristic times (indicated with vertical dashed lines) that will be used later.

Next, let us discuss the effect of the flatness criteria in the measurement of the error. Figure 2 shows the behavior of the error as a function of the modification factor $\ln f$, for increasing values of the flatness criteria for the WL algorithm. From this Figure, it is clear that no matter how flat the histogram is, the error always reaches a constant values, i.e., it saturates, even for very high value of the flatness criterion ($99.9 \%$).  

Figure 3, shows the behavior of the modification factor $F(t)$ as a function of time for two flatness criteria (Fig. 3a), and the corresponding error in the calculation of the critical temperature (Fig. 3b). Note that for $t<t_x$, the error curve corresponding to WL $80 \%$ is below the corresponding to WL $90 \%$. After that, for $t\geq t_x$, the behavior is reversed, that is, the curve corresponding to WL $80 \%$ is above the corresponding to WL $90 \%$. This can lead to an erroneous evaluation of the accuracy and precision. In fact, if the error is calculated at $t<t_x$, it is found that the error of WL $90 \%$ is greater than that of WL $80 \%$; however, if $t>t_x$, the behavior is the opposite.

Next, we discuss the range of validity of the conjecture of Zhou and Batt \cite{Zho05}, which assumes that the error is proportional to $\sqrt{\ln f_k}$, i.e., for a fixed value of $f_k$, the error will be the same for any flatness criteria. 

To visualize this, the value of $f_k$ is fixed in Fig. 3a (as example, $k=13$ which corresponds to $\ln f_k=1.2208 \times 10^{-4}$, horizontal dot line), in such a way that the intersection between the horizontal line and  the $F(t)$ curves with  $80 \%$ and $90 \%$ occurs at times $t_A$ and $t_B$, respectively. These times are less than $t_x$. The errors corresponding to these times, in the Fig. 3b, are the same, confirming the conjecture. However, for $t>t_x$ (for example, $t_C$ in the Fig 3a), the conjecture cannot be applied because of the saturation of the error in the WL algorithm. Then, one can say that, for the WL algorithm, the error is proportional to $\sqrt{\ln f_k}$, provided that the time $t<t_x$; in other words, the conjecture of Zhou and Bhatt is valid for $ \ln f_k \geq 1 / t $. In contrast, for the $1/t$ algorithm, the Zhou and Batt conjecture is valid for all time.

From the above the determination of $t_x$ is of fundamental importance, and it cannot be obtained using the WL algorithm, but using the $1/t$ algorithm instead.

\begin{figure}
\includegraphics[scale=0.75]{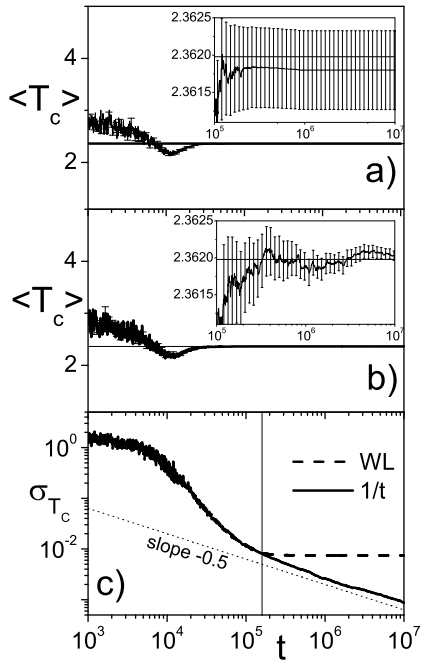}% Here is how to import EPS art
\caption{\label{fig:epsart} Behavior of the mean value of the critical temperature as a function of time and the confidence interval for a) the WL algorithm and b) the $1/t$ algorithm (a magnification of the curve is shown in the inset). In c), the behavior of the standard deviation for both algorithms is shown. The fitting of the $1/t$ curve gives an slope of $0.493(2)$.  The data represent the average of 200 independent realizations ($p$=200). }
\end{figure}

In order to compare the statistical and systematic errors for a single walker, one proceeds to calculate the mean value of the observable over $p$ independent realizations using Eq.(10) and  the corresponding  standard deviation $\sigma_{p}$ using Eq.(11) (remember that $m=1$).

Figure 4 shows the behavior of the mean value of the critical temperature as a function of time and the confidence interval ($\langle T_c \rangle \pm \frac{\sigma_{p}}{\sqrt{p}}$), for both WL (Fig. 4a) and $1/t$ (Fig. 4b) algorithms. In the inset of the Figure 4a and 4b, a magnification of the curves is shown. As observed, the standard deviation which is a measurement of the statistical error presents different behavior according to the algorithm used. In fact, $\sigma$ for WL algorithm,  decreases for $t<t_x$, after that, for $t>t_x$ it reaches a constant value. On the contrary,  $\sigma$  for $1/t$ algorithm, always decreases, and for $t>t_x$ decays as  $ 1/\sqrt{t}$, as shown in Figure 4c. The statistical error decreases with time for the $1/t$ algorithm, while it remains constant for the WL algorithm. It is important to note that for the WL algorithm, both the error measured by eqs. (4) and (9)  (Fig. 1b), as well as the standard deviation, eq.(11)  (Fig. 4c), reach a constant value for $t> t_x$. 
The unusual behavior of the standard deviation as a function of time has been discussed in reference \cite{Bela14}.

\begin{figure}
\includegraphics[scale=0.8]{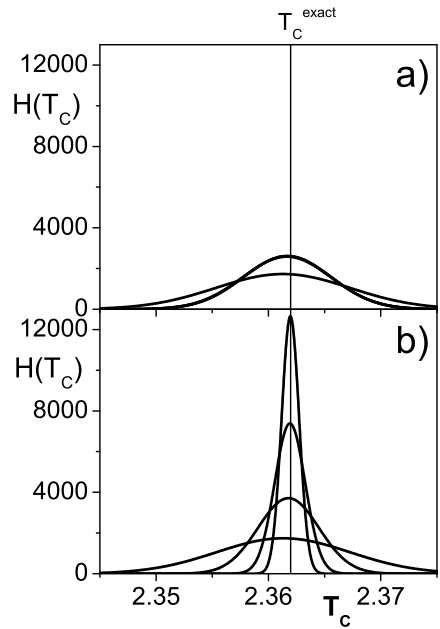}% Here is how to import EPS art
\caption{\label{fig:epsart} Best-fit Gaussians for the histograms of the critical temperatures obtained using: a) WL with  $80 \%$-flatness criterion and b) 1/t algorithms. Each of the curves corresponds to $p$=100000 independent runs. The curves are ordered from bottom to top according to the times defined in Fig.(1). Note that for the WL case, the curves collapse into each other for $t>t_x$.}
\end{figure}

\begin{figure}
\includegraphics[scale=0.8]{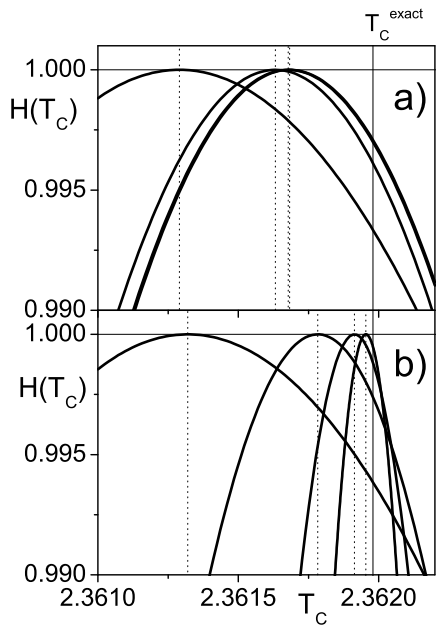}% Here is how to import EPS art
\caption{\label{fig:epsart}Magnification of the curves described in Figure 5. All the curves are normalized to one, for comparison purposes.}
\end{figure}

To confirm this effect, figure 5 shows the best-fit Gaussian for the histograms of the critical temperature obtained at times $t_2, t_3, t_4, t_5$ which are defined in Fig. 1, for the sampling using: a) WL using the $80\%$-flatness criterion and b) the $1/t$ algorithm; each histogram is obtained for  $p=$100000 independent runs. The vertical line corresponds to the exact temperature obtained with data from Ref.\cite{Beale96}. For comparing properly, the scales are the same in both Figures.

In Figure 6a and 6b, an enlargement of the curves is shown. In order to compare them, the curves are adequately normalized.  

It is observed that for $t>t_x$ ($t_3,t_4,t_5$), the curves corresponding to WL are superimposed (Fig. 5a, 6a), which is in agreement with the discussion above, i.e, the standard deviation is constant. On the contrary, for the $1/t$ algorithm, the standard deviation decreases with time (Fig. 5b, 6b). Note that at $t=t_2$, the curves coincide within statistical error. This is due to that this takes place in stage a) of the $1/t$ algorithm, which coincides with the WL algorithm.

In the rest of the this Section, the dependence of the errors as a function of the number of walkers $m$, for a fixed value of time $t$, is discussed. 

To get the DOS and the observables using the so-called $m$-random walkers algorithm, one proceeds as follows: a) the running time is fixed to a certain value, $t=t'$, then, $S(E,t=t')=\ln g(E,t=t')$ is obtained for all values of $E$; b) the algorithm is executed by $m$ independent random walkers; c) the  quantities of interest are averaged adequately. 

\begin{figure}
\includegraphics[scale=0.75]{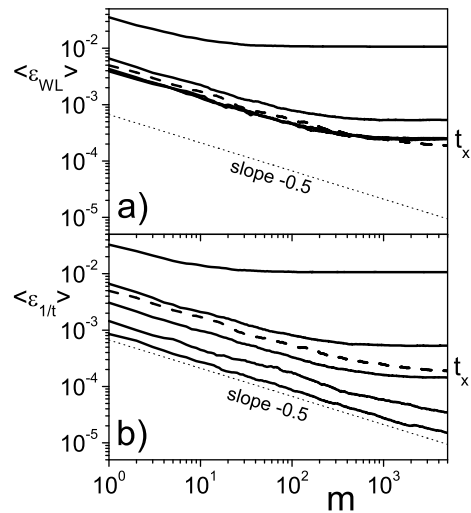}% Here is how to import EPS art
\caption{\label{fig:epsart} Behavior  of the error as a function of the number of walkers for the calculation of the DOS, using the WL algorithm (Fig. 7a) and the $1/t$ algorithm (Fig. 7b).  In both figures, the curves are shown in decreasing order according to the times $t_1, t_2, t_x, t_3, t_4$ and $t_5$, which are defined in Fig. 1. The dotted line with slope $1/\sqrt{m}$ is shown for comparison. The data represent the average of 100 independent realizations ($p=$100).}
\end{figure}

The error in the calculation of the DOS as a function of the number of walkers, using WL and $1/t$ algorithms, is shown in Figure 7a and 7b, respectively, where $t'$ takes the following values:  $t_1=3 \times 10^4$  MCS, $t_2=1 \times 10^5 $ MCS, $t_x=1.4 \times 10^5 $ MCS, $t_3=3 \times 10^5$ MCS, $t_4=1\times 10^6$ MCS and $t_5=3 \times 10^6$ MCS (indicated by vertical dotted lines in Figure 1).

As observed, the error decreases with the number of walkers as $1/\sqrt{m}$, and for a certain value of $m$, loses this functionality, approaching a constant value, i.e. the error is saturated with the number of walkers.

However, this behavior presents different characteristics according to the algorithms used.

For $t<t_x$, the $1/t$ algorithm is still in the WL regime (see figure 1). Therefore, it should be expected that the errors are statistically the same. In fact, this is confirmed in Figure 7a and 7b, where the error curves corresponding to $t_1=3 \times 10^4$ MCS and $t_2=1 \times 10^5 $ MCS (the first two curves from top to bottom) are the same. The error decreases with the number of walkers as $1/\sqrt{m}$ for $ m \leq 100$ for the top curve, and for $m\leq 400$ for the next curve; then it loses this functionality, approaching a constant value. In other words, there is a critical number $m_x$ that separates this two regimes, such that, for $m<m_x$, the error goes as $1/\sqrt{m}$, and for $m\geq m_x$, the error reaches a constant value (saturation value).

\begin{figure}
\includegraphics[scale=0.75]{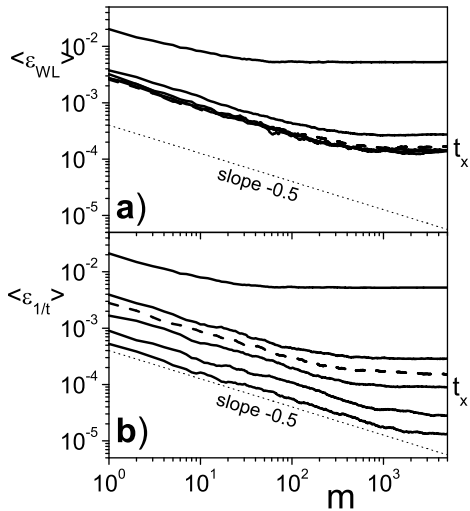}% Here is how to import EPS art
\caption{\label{fig:epsart} Behavior of the error as a function of the number of walkers for the calculation of the critical temperature using the WL algorithm (Fig. 8a) and the $1/t$ algorithm (Fig. 8b).  The parameters are the same as in Figure 7. The data represent the average of 100 independent realizations ($p=$100).}
\end{figure}

\begin{figure}
\includegraphics[scale=0.75]{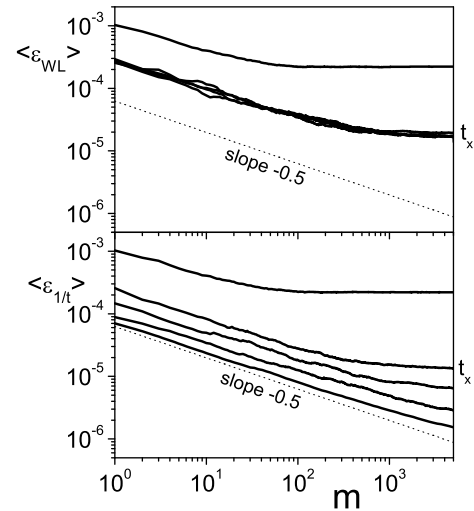}% Here is how to import EPS art
\caption{\label{fig:epsart} Behavior of the error as a function of the number of the walkers for the calculation of number $\pi$, using the WL algorithm (Fig. 9a) and the $1/t$ algorithm (Fig. 9b).  In both figures, the curves are shown in decreasing order according to the times $t_1, t_2, t_x, t_3, t_4$ and $t_5$, defined in Fig. 1. The parameters are the same as in Fig. 7. The data represent the average of 100 independent realizations ($p=$100).}
\end{figure}

Although this behavior  is observed in all cases, the error calculated by the WL algorithm for $t\geq t_x$ has a peculiar characteristic, which is that all the error curves collapse into a single curve (see Figure 7a). That means that, for $t\geq t_x$, $m_x$ is the same for all curves; while, in the $1/t$ algorithm, the error curves do not collapse (see Figure 7b); and therefore $m_x$ increases with time.

It is important to note that the WL algorithm does not determine the saturation time, $t_x$, which is critical when running the algorithm properly, since for longer times, it becomes an unnecessary calculation. Therefore the WL algorithm, could be inefficient, since for $t \geq t_x$, the error curves collapse into one, regardless of the number of walkers used. For example, in this particular system (two-dimensional Ising Model with size $L=8$, periodic boundary conditions and using $80 \%$-flatness criterion), the saturation time is $t_x \approx 140000$ MCS and the critical number of walkers is $m_x\approx 400$. By simple inspection of the figure, it seems to be impossible to obtain an error below $10^{-4}$ in the calculation of the DOS using the WL algorithm, either by increasing the running time ($t>t_x$) or the number of walkers ($m>m_x$).

\begin{figure}
\includegraphics[scale=0.8]{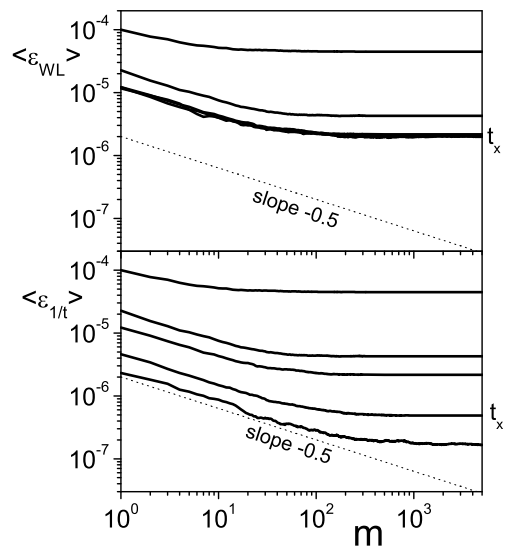}% Here is how to import EPS art
\caption{\label{fig:epsart} Behavior of the error as a function of the number of walkers for the calculation of the DOS for a window of $ 300 $ energy sites,  corresponding to a two-dimensional Ising Model with size $L=64$, using the WL algorithm (Fig. 10a) and the $1/t$ algorithm (Fig. 10b). The data represent the average of 100 independent realizations ($p=$100).}
\end{figure}

This behaviour is also observed in the calculation of the critical temperature $T_c$ with the number of walkers $m$ (see Figure 8a and 8b), and in the continuum approximation, i.e., the multidimensional numerical integration to calculate the number $\pi$ (see Figure 9a and 9b).

The effect of the size of the system in the behavior of error in the calculation of the DOS, can be important for real systems. 
However, it is expected that the characteristics described above to be the same as for small systems. That is, if there is a number of walkers $ m_x $ to which the error saturates  for a small system, this must occur to a larger system. To confirm this, shown in Figure 10 as is the behavior of the error in the calculation of the DOS as a function of $m$ for a window of $ N=300 $ energy sites,  corresponding to a two-dimensional Ising Model with size $L=64$. The energy range is between $(-0.29, 0]$ (energy per sites). This is a usual size of window used in this particular case. As shown, the general behavior is the same as the described in the previous cases. 

\begin{figure}
\includegraphics[scale=0.8]{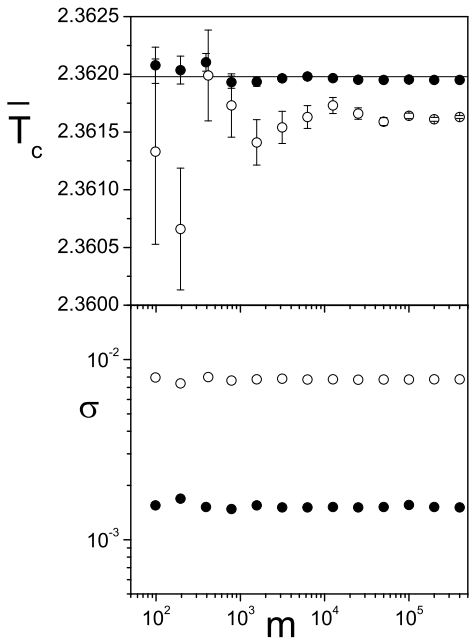}% Here is how to import EPS art
\caption{\label{fig:epsart} a)  Behavior of the mean value of the critical temperature, $\overline{T_c}$, as a function of the number of walkers $m$ for the WL algorithm and the $1/t$ algorithm, calculated at fixed time $t=t_5$; the confidence interval is also shown for both curves. b) Behavior of the standard deviation as a function of $m$ for both WL and $1/t$ algorithm. Filled (empty) symbols represent $1/t$ (WL) procedures, respectively. }
\end{figure}

\begin{figure}
\includegraphics[scale=0.8]{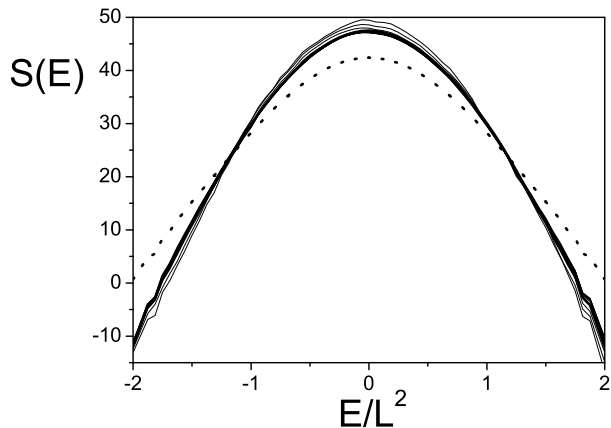}% Here is how to import EPS art
\caption{\label{fig:epsart} Density of state (DOS) for a fixed value of $t'=10000$ MCS, and different values of $m$ ($m$=10,20, 30, 40,...200, 2000, from top to bottom in $E/L^2=0$ ) . In this case the critical value of the number of walker is $m_x=100$. The exact value of the DOS is denoted by point line .}
\end{figure}

Figure 11a shows the behavior of the mean value of the critical temperature $\overline{T_c}$ as a function of $m$, at a fixed time $t_5=3\times 10^6$ MCS, using WL (open symbols) and $1/t$ (filled symbols) algorithms; the error bars ($\overline{T_c} \pm \frac{\sigma_{m}}{\sqrt{m}}$ ) are also shown. The data represent one realization ($p=$1). The exact value of the critical temperature $T_c^{exc}$, is indicated with a horizontal line. 

The mean value calculated by the WL algorithm is further away from the exact value of $T_c$, than the calculated by the $1/t$ algorithm. The standard deviation remains constant in both cases. The value of $\sigma$ for the WL algorithm is always greater than the corresponding to the $ 1 / t $ algorithm; this is confirmed in Figure 11b. Looking at Figure 11a, one can observed that the systematic error, which is a measure of the distance between the average value of the calculated critical temperature and the exact one, is greater than the statistical error for increasing values of $m$, for both algorithms.

With respect to understand the reason for the saturation of the DOS to the number of walkers, in Figure 12, the behavior of the normalized DOS is plotted as a function of $E/L^2$. Note that the DOS is normalized to the mean value.
Different values of $m$ are used (in the Figure, the values of $m$ increase from top to bottom at $E/L^2=0$) at fixed time $t'=10000$.  Note that the DOS approach to some limiting value $S(t)=S_{lim} (t')$ (thick black line), which differs from the exact one (dotted line). 
It is interesting to note that, the growth of the DOS is skewed,  because it overestimates the most likely energy configurations (central part of the energy range) and underestimates the less likely energy  configurations (left and right parts of the energy range). If one makes measurements DOS as a function of $ m $, the value of $ S (t', m) $ will approach $ S_{lim} (t')$ and not the exact value, as $ 1 / \sqrt { m} $ (thin black line in Figure 12).
For $m>m_x$,  the mean value will not be altered by new measurements, because they will not differ from it. This is, the mean value is more accurate, but it differs from the exact one, that is, the exactness does not change for $m>m_x$.  

For $t'<t_x$ the  $S_{lim}(t',m)$ changes with time for both algorithms. However, for $t'>t_x$ the behavior is very different. In fact, for the WL algorithm, $S_{lim}(t',m)$  does not change in time (it is frozen). Then, no matter the number of walkers used to calculate it, the mean value is also saturated.  In other words, $m_x$ does not change, and the error curves collapse into a single one (see Figs. 7a). 

On the contrary,  for the $1/t$ algorithm, for $t'>t_x$, $S_{lim}(t',m)$  approaches asymptotically to the exact value. The longer the time, the closer to the exact value. So, $m_x$ changes with the value of $t'$.

From the above, it is clear that by increasing the running time, the $1/t$ measurement can be improved, but not the WL, since for $t>t_x$, any measurement will saturate. 
On the other hand, it makes no sense to increases the number of walkers above a critical value $m_x$, since it does not reduces the error in the calculation. Therefore the number of walkers does not guarantee convergence. 

Summarizing, for a single random walker one can writ: 
\begin{eqnarray}
\lim_{t \to \infty} \varepsilon_{WL}(t) \rightarrow const
\end{eqnarray}
for the WL algorithm,
\begin{eqnarray}
\lim_{t \to \infty} \varepsilon_{1/t}(t) \rightarrow 0;
\end{eqnarray}
 for the $1/t$ algorithm; while for $m$-random walkers,
\begin{eqnarray}
\lim_{m \to \infty} \varepsilon_{WL}(t,m)|_{t=t'} \rightarrow C_1
\end{eqnarray}
and for the $1/t$ algorithm, 
\begin{eqnarray}
\lim_{m \to \infty} \varepsilon_{1/t}(t,m)\bigg |_{t=t'} \rightarrow C_2 
\end{eqnarray}
where $C_1$ and $C_2$ are constants, which fulfill the following conditions: for $t<t_x$, $C_1=C_2$, while for $t>t_x$, $C_1>C_2$.

\section{CONCLUSIONS}

Before presenting the conclusions about the convergence problem of the m-random walkers, it is convenient to revisit the convergence properties of a single random walker. It has been analytically demonstrated that the exponential decrease of the modification factor,  $F=\ln f$, with the number of iterations, is the reason for the saturation of the error in WL the algorithm with a single random walker.  

One important conclusion that is not mentioned before is that for the WL algorithm, the error is proportional to $\sqrt{\ln f_k}$, provided that the time $t<t_x$; in other words, the conjecture of Zhou and Bhatt is only valid for $ \ln f_k \geq 1 / t $.  By contrast, for the $1/t$ algorithm, the conjecture is valid for all time. 

An interesting feature of the single random walker is the comparison between the statistical and systematic errors. The first one is proportional to the standard deviation, while the second one is proportional to the distance between the mean value and the exact one. As discussed in the text, one can observe that the standard deviation for the WL algorithm for $t>t_x$ remains constant, while for the $1/t$ algorithm it decreases as  $ 1/\sqrt{t}$. Thus, the statistical error for the WL algorithm is greater than the corresponding to the $1/t$ algorithm. On the other hand, systematic error corresponding to the WL algorithm remains constant, while for $1/t$ algorithm it decreases as $1/\sqrt{t}$ (see Fig. 1b). 

The building of the density of states, and consequently, the calculation of different observables, by using entropic sampling methods (WL and $1/t$- algorithms) with multiple random walkers, have convergence problems.
In fact, if the error is calculated using $m$ experiments (walkers) at a fixed time, $t$, it decreases with the number of walkers as $ 1/\sqrt{m}$ until reaching a certain value of walkers $m=m_x$ from which it saturates. This critical number $m_x$ separates these regimes, such that, for $m<m_x$, the error goes as  $1/\sqrt{m}$, and for $m\geq m_x$, the error reaches a constant value (saturation value). The critical value $m_x$ depends on the characteristics of the system (size, interactions, connectivity, etc.).

The saturation of the error with the number of walkers is observed in the WL as well as in the $1/t$. However, there are substantial differences for both algorithms.

As observed, the critical value $m_x$ is lower in the WL than in the $1/t$. Moreover, the saturation for the $1/t$ occurs at very high number of walkers. On the other hand, the WL algorithm presents a peculiar behavior, that is, for $t\geq t_x$, all the error curves calculated by the $m$-walkers collapse into a single curve; while in the case of the $1/t$ algorithm, the error curves do not; therefore, it makes no sense to run the WL algorithm for $t > t_x$.

It is shown that the statistical error is reduced with the number of walkers. However, the systematic error depends on the algorithm. It is also shown that WL algorithm presents a systematic error that is not reduced with the number of walkers; the precision increases but not the exactness.

For a given system, the WL algorithm cannot calculate the DOS with greater accuracy than a certain value, even when the time is increased ($t$ tends to infinity) or the number of walkers is increased ($m$ tends to infinity). However, with the implementation of the $1/t$ algorithm, the error will be always reduced because there is no saturation in time.

Summarizing, one can claim that the $1/t$ algorithm is convergent. That is, the error in calculating the density of states versus time, tends to zero as $t$ tends to infinity. This was previously demonstrated numerically \cite{Bela07a} and analytically [\onlinecite{Bela07b,Bela14}]. In contrast, the WL algorithm is not convergent, i.e., the error saturates at a finite time. 

The  $1/t$ algorithm is always more efficient than the WL algorithm, even when it is as a function of the number of walkers. However, to calculate the DOS with high accuracy by using the $1/t$ algorithm, it is better to run the algorithm as a function of time, and not as a function of the number of walkers.

In conclusion, it makes no sense to increase the number of parallel programs (number of walkers), above a critical value $m_x$, since it does not reduce the error in the calculation. The number of walkers does not guarantee convergence.

The authors would like to thank Prof. Dr. A. P. Velasco for the useful discussion and Dr. O. J. Furlong for reading the manuscript. This work is partially supported by the CONICET (Argentina).


\begin{thebibliography}{99}
\bibitem{Flatt00} In this paper, the flatness criteria means that, histogram $H(j)$ for all possible $j$ is not less than $80$ per cent of $\langle H \rangle=\frac{1}{L}\sum_{i=1}^{L}H(i)$ \cite{Wang01b}.
\end{thebibliography}
\end{document}